\documentclass[twocolumn,pra,aps,showpacs]{revtex4}
\usepackage{amsmath}
\usepackage{epsfig}

\begin{document}

\title{Landau-Zener Tunnelling in a Nonlinear Three-level System}
\author{Guan-Fang Wang $^{1,2}$, Di-Fa Ye$^1$, Li-Bin Fu$^1$, Xu-Zong Chen$^3$, and Jie Liu$^{1*}$}
\affiliation{$^1$Institute of Applied Physics and Computational
Mathematics, P.O. Box 8009 (28), 100088 Beijing, China}
\affiliation{$^2$Institute of Physical Science and Technology,
Lanzhou University, 730000 Lanzhou, China} \affiliation{$^3$ Key
Laboratory for Quantum Information and Measurements, Ministry of
Education, School of Electronics Engineering and Computer Science,
Peking University, Beijing 100871, China}

\begin{abstract}
We present a comprehensive analysis of the Landau-Zener tunnelling
of a nonlinear three-level system in a linearly sweeping external
field. We find the presence of  nonzero tunnelling probability in
the adiabatic limit (i.e., very slowly sweeping field) even for
the situation that the nonlinear term is very small and the energy
levels keep the same topological structure as that of linear case.
In particular, the tunnelling is irregular with showing an
unresolved sensitivity on the sweeping rate. For the case of
fast-sweeping fields, we derive an analytic expression for the
tunnelling probability with stationary phase approximation and
show that the nonlinearity can dramatically influence the
tunnelling probability when the nonlinear "internal field"
resonate with the external field.  We also discuss the asymmetry
of the tunnelling probability induced by the nonlinearity. Physics
behind the above phenomena is revealed and possible application of
our model to triple-well trapped Bose-Einstein condensate is
discussed.
\end{abstract}

\pacs{03.75.-b, 05.45.-a, 03.75.Kk, 42.50.Vk } \maketitle

\section{Introduction}

Avoided crossing of energy levels is a universal phenomenon for the quantum
non-integrable systems where the symmetry break leads to the splitting of
degenerate energy levels forming a tiny energy gap. Around the avoided
crossing point of the two levels the Landau-Zener tunnelling (LZT) model
provides an effective description for the tunnelling dynamics under
assumption that the energy bias of two levels undergoes a linear change with
time\cite{landau}. It is a basic model in quantum mechanics and has
versatile applications in quantum chemistry \cite{may}, collision theory %
\cite{harmin}, and more recently in the spin tunnelling of nanomagnets \cite%
{wensdorf1,wensdorf2}, Bose-Einstein condensates (BEC) \cite{yurovsky} and
quantum computing \cite{shytov0}, to name only a few.

LZT model has been extended to many versions taking diverse
physical conditions into account: LZT problem with a time-varied sweeping rate%
\cite{garanin}, LZT model with a fast noise from the  outer
environment \cite{pok}, LZT model with periodic modulation
\cite{duan,wang}, and so on. Among them,
LZT in a nonlinear two-level system is one of most interesting models
and attracts much attention recently\cite%
{raghavan,bwu,zobay,jliu1}. In this model, the level energies
depend on the occupation of the levels, may arise in a meanfield
treatment of a many-body system where the particles predominantly
occupy two energy levels. The nonlinear LZT model not only
demonstrate many novel behavior of great interest in theory but
also has important applications in spin tunnelling of nanomagnets
\cite{jliu} and a Bose-Einstein condensate in a double-well
potential \cite{raghavan,zobay,michael} or in an optical lattice
\cite{bwu,jliu1}. However, since most of the problems of interests
involve more than two energy levels, with transitions between
several levels happening
simultaneously\cite{shytov,valentine,vitanov,carroll}, for
example, BECs trapped in multiple
wells\cite{nemoto,roberto,sun,graefe}, spin tunnelling of
nanomagnets with large spin, etc. It is naturally desirable to
extend the above nonlinear tunnelling to the multi-level
situation.

In present paper, we consider the simplest multi-level
system---three-level system, to investigate its complicated
tunnelling dynamics in the presence of  nonlinearity. Because
quantum transitions may happen between several levels
simultaneously, the LZT in  the nonlinear three-level model show
many striking properties distinguished from that of the two-level
case. In the adiabatic limit we will show that, for a very small
nonlinear parameter that the energy levels still keep the same
topological structure as its linear counterpart, the adiabaticity
breaks down manifesting the presence of a nonzero tunnelling
probability. This is quite different from the two-level case,
where the break down of the adiabaticity is certainly accompanied
by a topological change on the energy levels. More interestingly,
the tunnelling is irregular with showing an unresolved sensitivity
on the sweeping rate, a phenomenon attributed to the existence of
chaotic state. In the sudden limit, we derive an analytic
expression for the tunnelling probability under stationary phase
approximation and show that the nonlinearity can dramatically
influence the tunnelling probability at the resonance between the
nonlinear "internal field" and the external field. We also discuss
the asymmetry of the tunnelling probability induced by the
nonlinearity. The physical mechanism behind these phenomena is
revealed and possible application of our model to triple-well
trapped Bose-Einstein condensate is discussed.

The paper is organized as follows. In Sec.II we introduce our
nonlinear three-level LZT model and calculate its adiabatic
levels. Section III discusses LZT among the levels. Section IV
gives a possible application of the model to the triple-well
trapped BEC.

\section{The model and adiabatic levels}

We consider following dimensionless Schr\"{o}dinger equation
\begin{equation}
i\frac{d}{dt}\left(
\begin{array}{c}
a_{1} \\
a_{2} \\
a_{3}%
\end{array}%
\right) =H\left(
\begin{array}{c}
a_{1} \\
a_{2} \\
a_{3}%
\end{array}%
\right)  \label{equation1}
\end{equation}
with the Hamiltonian given by
\begin{equation}
H=\left(
\begin{array}{ccc}
\frac{\gamma }{2}+\frac{c}{4} \left| a_{1}\right| ^{2} & -\frac{v}{2} & 0 \\
-\frac{v}{2} & \frac{c}{4} \left| a_{2}\right| ^{2} & -\frac{v}{2} \\
0 & -\frac{v}{2} & -\frac{\gamma}{2}+\frac{c}{4} \left| a_{3}\right| ^{2}%
\end{array}%
\right)  \label{euqation2}
\end{equation}%
where $v$ is the coupling constant between the neighboring levels; $c$ is
the nonlinear parameter; the energy bias $\gamma$ is supposed to be adjusted
by a linearly external filed, i.e., $\gamma =\alpha t$, $\alpha$ is the
sweeping rate; $a_{1}$, $a_{2}$, $a_{3}$ is probability amplitude in each
level and the total probability $\left| a_{1}\right| ^{2}+\left|
a_{2}\right| ^{2}+\left| a_{3}\right| ^{2}$ is conserved and set to be unit.

When the nonlinear parameter vanishes, our model reduces to the linear case
and the adiabatic energy levels $\varepsilon (\gamma )=0,\pm \frac{1}{2}%
\sqrt{\gamma^{2}+2v^{2}}$ (Fig.1(a)) derived by diagonalizing the
Hamiltonian (2). Tunnelling probability $\Gamma _{nm}$
$(n,m=1,2,3)$ is defined as the occupation probability on the m-th
level at $\gamma \rightarrow +\infty$ for the state initially on
the n-th level at $\gamma \rightarrow -\infty$. For the linear
case, the above system is solvable analytically and the tunnelling
probabilities can be explicitly expressed as \cite{carroll}
\begin{equation}
\Gamma _{11}=\left[ 1-\exp \left( -\frac{\pi v^{2}}{2\alpha }\right) \right]
^{2}  \label{euqation3}
\end{equation}
\begin{equation}
\Gamma _{12}=2\exp (-\frac{\pi v^{2}}{2\alpha })\left[ 1-\exp \left( -\frac{%
\pi v^{2}}{2\alpha }\right) \right]  \label{euqation4}
\end{equation}
\begin{equation}
\Gamma _{13}=\exp (-\frac{\pi v^{2}}{\alpha })  \label{euqation5}
\end{equation}
\begin{equation}
\Gamma _{22}=\left[ 1-2\exp \left( -\frac{\pi v^{2}}{2\alpha }\right) \right]
^{2}  \label{euqation6}
\end{equation}
The others are $\Gamma _{21}=\Gamma _{23}=\Gamma _{32}=\Gamma _{12}, \Gamma
_{31}=\Gamma _{13}, \Gamma _{33}=\Gamma _{11}$ due to the symmetry of the
levels.

With the presence of the nonlinear terms, we want to know how the tunnelling
dynamics in the above system is affected. In our discussions, the coupling
parameter is set to be unit, i.e., $v=1$. Therefore, weak nonlinear case and
strong nonlinear case mean that $c <<1$ and $c>>1$, respectively. As to the
external fields, we will consider three cases, namely, adiabatic limit,
sudden limit, and moderate case, corresponding to $\alpha <<1$, $\alpha >>1$
and $\alpha \sim 1$, respectively.

%%%%%%%%%%%%%%%%%%%%%%%%%%%%%%%%%%%%%%%%%%%%%%%%%%%%%%%%%%%%%%%%%%%%%%%%%%%%%%%%%%%%%%%%%
\begin{figure}[tbh]
\begin{center}
\rotatebox{0}{\resizebox *{9.0cm}{11.0cm} {\includegraphics
{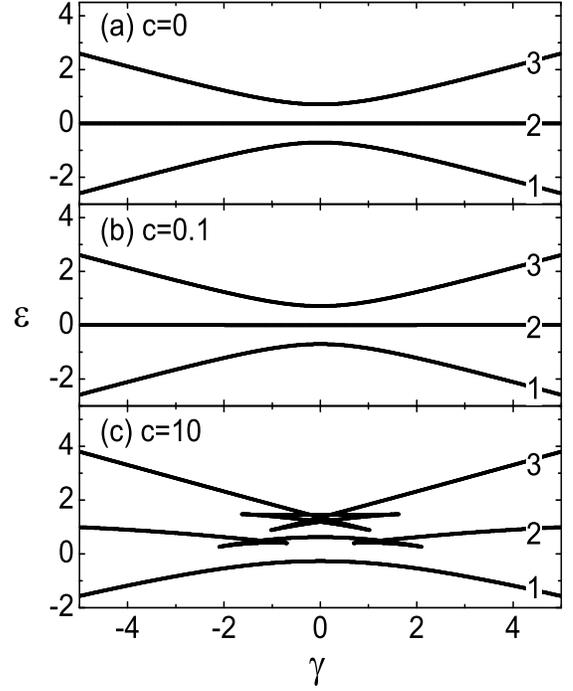}}}
\end{center}
\caption{Adiabatic three levels at $v=1.0$. (a) linear case, (b)
weak nonlinearity case of $c=0.1$. (c) strong nonlinearity case of
$c=10.0$.} \label{fig.1}
\end{figure}

%%%%%%%%%%%%%%%%%%%%%%%%%%%%%%%%%%%%%%%%%%%%%%%%%%%%%%%%%%%%%%%%%%%%%%%%%%%%%%%%%%%%%%%%%%

Similar to the linear case, we need to analyze  the adiabatic
levels of the nonlinear model first. With
$a_{1}=\sqrt{s_{1}}e^{i\theta
_{a_{1}}},a_{2}=\sqrt{1-s_{1}-s_{2}}e^{i\theta
_{a_{2}}},a_{3}=\sqrt{s_{2}}e^{i\theta _{a_{3}}}$, we introduce
the relative phase $\theta _{1}=\theta _{a_{1}}-\theta _{a_{2}},
\theta _{2}=\theta _{a_{3}}-\theta _{a_{2}}$. In terms of
$s_{1},\theta _{1}$ and $s_{2},\theta _{2}$, the nonlinear
three-level system is casted into a classical Hamiltonian system,
$$
H_{e}=\left( \frac{\gamma }{2}+\frac{c}{8}s_{1}\right) s_{1}+\frac{c}{8}%
\left( 1-s_{1}-s_{2}\right) ^{2} +\left( -\frac{\gamma }{2}+\frac{c}{8}%
s_{2}\right) s_{2}
$$
\begin{equation}
-v\sqrt{(1-s_{1}-s_{2})s_{1}}\cos \theta _{1}-v\sqrt{(1-s_{1}-s_{2})s_{2}}%
\cos \theta _{2}
\end{equation}
$s_{1},\theta _{1}$ and $s_{2},\theta _{2}$ are two pairs of
canonically conjugate variables of the classical Hamiltonian
system. The fixed points of the nonlinear classical Hamiltonian
correspond to the eigenstates of the nonlinear three-level system,
and are given by the following equations:

\begin{equation}
\dot{s_{1}}=-v\sqrt{(1-s_{1}-s_{2})s_{1}}\sin \theta _{1}  \label{euqation7}
\end{equation}%
\begin{eqnarray}
\dot{\theta _{1}} &=&\frac{\gamma }{2}-\frac{c}{4}(1-2s_{1}-s_{2})-\frac{%
1-2s_{1}-s_{2}}{2\sqrt{(1-s_{1}-s_{2})s_{1}}}v\cos \theta _{1}  \notag \\
&&+\frac{s_{2}}{2\sqrt{(1-s_{1}-s_{2})s_{2}}}v\cos \theta _{2}
\end{eqnarray}%
\begin{equation}
\dot{s_{2}}=-v\sqrt{(1-s_{1}-s_{2})s_{2}}\sin \theta _{2}  \label{euqation8}
\end{equation}%
\begin{eqnarray}
\dot{\theta _{2}} &=&-\frac{\gamma }{2}-\frac{c}{4}(1-s_{1}-2s_{2})+\frac{%
s_{1}}{2\sqrt{(1-s_{1}-s_{2})s_{1}}}v\cos \theta _{1}  \notag \\
&&-\frac{1-s_{1}-2s_{2}}{2\sqrt{(1-s_{1}-s_{2})s_{2}}}v\cos \theta _{2}
\label{equation 10}
\end{eqnarray}%
By solving the equations (\ref{euqation7})-(\ref{equation 10}) the
eigenstates of the system are obtained. Accordingly,  the
eigenenergy is obtained by $\epsilon = H_e$ i.e., the energy
levels are gained as shown in Fig.1.

For weak nonlinearity, the levels' structure is similar to its
linear counterpart (fig.1(b)). For  strong nonlinearity
(fig.1(c)), in the mid-level a double-loop topological structure
emerges and in the upper-level a butterfly structure appears.
Because of these  topological distortions on the energy levels, we
expect that  the tunnelling dynamics will dramatically change.

\section{Landau-Zener tunnelling}

In this section we study LZT in the nonlinear three-level system
both numerically and analytically. First, we consider two limit
cases: adiabatic limit, sudden limit, respectively. Then we will
discuss the tunnelling probability in general case and investigate
the symmetry of the tunnelling probability.

\subsection{Adiabatic limit ($\protect\alpha\ll 1$)}

In adiabatic limit, the characters of the tunnelling probabilities
should be  entirely determined by the topology of the energy
levels and the eigenstates' properties (corresponding to the
stability of the fixed points in classical Hamiltonian system),
according to the adiabatic theorem\cite{jliu2,fupre}. So, we
expect that, for the weak nonlinearity case, an initial state
started from  any levels (upper, mid or lower) will follow the
levels and evolves adiabatically, as a result,  no quantum
transition between levels occurs; for the strong nonlinearity, an
initial state from the lower level is expected to evolve
adiabatically keeping stay on the ground state, leading to zero
adiabatic tunnelling probability, whereas for the state initially
from the mid or upper level, due to the topological change of the
level, it can not move smoothly from left-side to the right-side.
Transition to other levels happens at the tip of the loop or
butterfly. Consequently, the adiabatic tunnelling probability is
expected to be nonzero.

However, the above picture is only partly corroborated by our
directly solving the Schr\"odinger equation using forth-fifth
order Runge-Kutta  adaptive-step algorithm, as shown in Fig.2.

%%%%%%%%%%%%%%%%%%%%%%%%%%%%%%%%%%%%%%%%%%%%%%%%%%%%%%%%%%%%%%%%%%%%%%%%%%%%%%%%%%%%%%%%%
\begin{figure}[tbh]
\begin{center}
\rotatebox{0}{\resizebox *{10.0cm}{10.0cm} {\includegraphics
{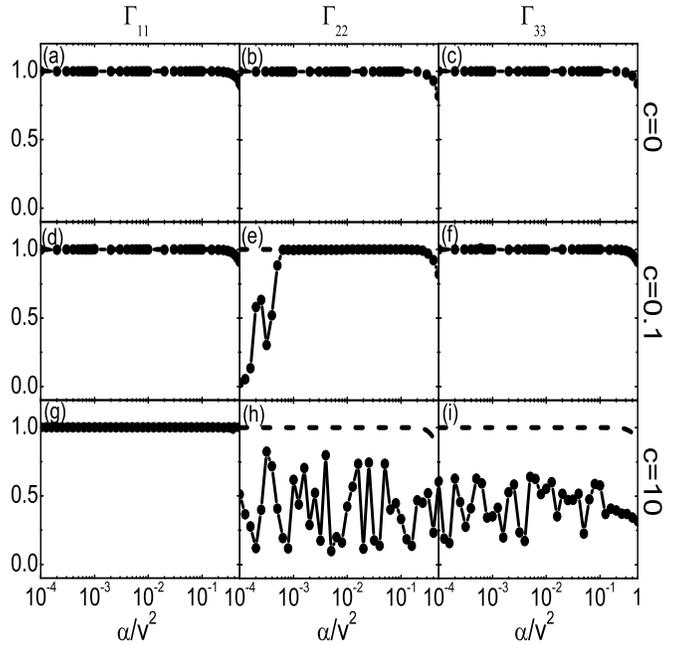}}}
\end{center}
\caption{The tunnelling probability $\Gamma _{11}$, $\Gamma
_{22}$, $\Gamma _{33}$ (full circles) as functions of
$\protect\alpha $ for different nonlinear parameter at $v=1.0$.
The dash lines represent the results from the linear Landau-Zener
model for comparison.} \label{fig.2}
\end{figure}
%%%%%%%%%%%%%%%%%%%%%%%%%%%%%%%%%%%%%%%%%%%%%%%%%%%%%%%%%%%%%%%%%%%%%%%%%%%%%%%%%%%%%%%%%%

On the one hand, Fig.2 clearly shows that, for the strong
nonlinearity case, as we expect, no tunnelling for the state from
the lower level, but a serious adiabatic tunnelling is observed
for the states from upper two levels. In particular, we find that
the tunnelling probability as a function of the sweeping rate
shows an irregular oscillation. We associate this irregularity to
the chaotic state. To demonstrate it, we plot in Fig.3 the
Poincare section of the trajectories for $c=10$ before and after
the tip of the butterfly structure of the upper level in Fig.1c.
It shows that, before the tip, the eigenstate corresponds to the
fixed point surrounded by quasi-periodic orbit, therefore is
stable. As the state evolves to the right tip of the butterfly, it
contact with chaotic sea, after that the state become chaotic. The
characteristics of the chaos is sensitive on the parameters,
therefore the chaotic state is responsible for  the irregular
tunnelling probability exposed by Fig.2h,i.

%%%%%%%%%%%%%%%%%%%%%%%%%%%%%%%%%%%%%%%%%%%%%%%%%%%%%%%%%%%%%%%%%%%%%%%%%%%%%%%%%%%%%%%%%
\begin{figure}[tbh]
\begin{center}
\rotatebox{0}{\resizebox *{8.0cm}{6.0cm} {\includegraphics
{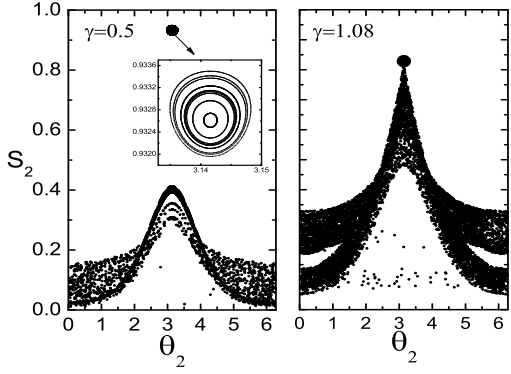}}}
\end{center}
\caption{Poincare section of the trajectories for $c=10$ before
and after the tip of the butterfly structure of the upper level in
Fig.1c.} \label{fig.3}
\end{figure}

%%%%%%%%%%%%%%%%%%%%%%%%%%%%%%%%%%%%%%%%%%%%%%%%%%%%%%%%%%%%%%%%%%%%%%%%%%%%%%%%%%%%%%%%%%
On the other hand, Fig.2 also shows that for the weak
nonlinearity, even though the adiabatic levels keeps the same
topological structure as the linear case, there is still  nonzero
tunnelling probability for the state started from the mid-level.
The tunnelling also shows some kind of irregularity. This
phenomenon counter to our naive conjecture from observing the
topological structure of the adiabatic levels.

To explain this unusual  phenomenon, we need make detailed
analysis on the  property of the fixed points of the classical
system Hamiltonian(7), corresponding to  the eigenstates of the
mid-level.

We plot quantity $s_{1}$ as the function of  $\gamma $ in
Fig.4 (a,b), we see the adiabatic evolution of the eigenstate breaks down around $%
\gamma =-2$ due to the nonlinearity (Fig.4 (b)). This adiabaticity
breakage is caused by the change on the property  of the fixed
point corresponding to the eigenstate of the mid-level. This is
revealed by investigating the Hamiltonian-Jaccobi matrix
obatined by linearizing the nonlinear equations (%
\ref{euqation7})-(\ref{equation 10}) at fixed points,
\begin{equation}
H_{J}=\left(
\begin{array}{cccc}
-\frac{\partial ^{2}H_{e}}{\partial s_{1}\partial \theta _{1}} & -\frac{%
\partial ^{2}H_{e}}{\partial ^{2}\theta _{1}} & -\frac{\partial ^{2}H_{e}}{%
\partial s_{2}\partial \theta _{1}} & -\frac{\partial ^{2}H_{e}}{\partial
\theta _{2}\partial \theta _{1}} \\
\frac{\partial ^{2}H_{e}}{\partial ^{2}s_{1}} & \frac{\partial ^{2}H_{e}}{%
\partial \theta _{1}\partial s_{1}} & \frac{\partial ^{2}H_{e}}{\partial
s_{2}\partial s_{1}} & \frac{\partial ^{2}H_{e}}{\partial \theta
_{2}\partial s_{1}} \\
-\frac{\partial ^{2}H_{e}}{\partial s_{1}\partial \theta _{2}} & -\frac{%
\partial ^{2}H_{e}}{\partial \theta _{1}\partial \theta _{2}} & -\frac{%
\partial ^{2}H_{e}}{\partial s_{2}\partial \theta _{2}} & -\frac{\partial
^{2}H_{e}}{\partial ^{2}\theta _{2}} \\
\frac{\partial ^{2}H_{e}}{\partial s_{1}\partial s_{2}} & \frac{\partial
^{2}H_{e}}{\partial \theta _{1}\partial s_{2}} & \frac{\partial ^{2}H_{e}}{%
\partial ^{2}s_{2}} & \frac{\partial ^{2}H_{e}}{\partial \theta _{2}\partial
s_{2}}%
\end{array}%
\right)  \label{euqation9}
\end{equation}%
We solve the eigenvalues of $H_{J}$ for different $\gamma $ and
plot our results in Fig.4c. These eigenvalues can be real, complex
or pure imaginary. Only pure imaginary eigenvalues corresponds to
the stable fixed point, others indicate the  unstable ones. In
Fig.4 (c),we can see the eigenvalues are complex number (i.e.,
their real parts are not zero) around $\gamma =0,\pm 2$. The
corresponding fixed points are unstable. For other regions, the
eigenvalues of $H_{J}$ are pure imaginary. Therefore, even though
no topological structure changes on the level structures, the
instability of the fixed point corresponding to the mid-level
leads to the breakdown of the adiabaticity manifesting the
irregular nonzero tunnelling  probability exposed by Fig.2e in the
adiabatic limit.

%%%%%%%%%%%%%%%%%%%%%%%%%%%%%%%%%%%%%%%%%%%%%%%%%%%%%%%%%%%%%%%%%%%%%%%%%%%%%%%%%%%%%%%%%
\begin{figure}[tbh]
\begin{center}
\rotatebox{0}{\resizebox *{8.0cm}{8.0cm} {\includegraphics
{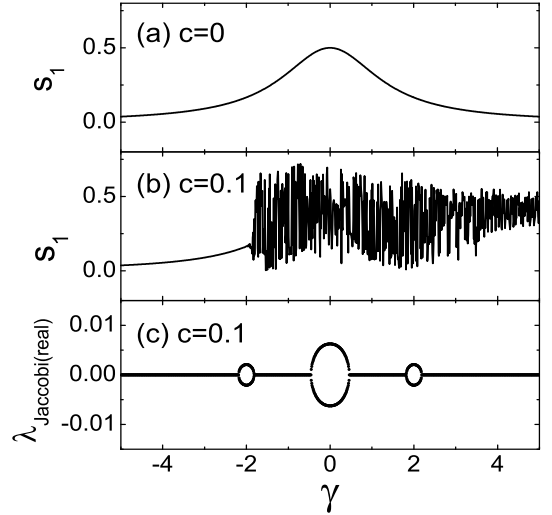}}}
\end{center}
\caption{The variety of $s_{1}$ with $\protect\gamma $ when the eigenstate $%
\left( 0,1,0\right) ^{T}$ evolve adiabatically at $v=1.0,\protect\alpha %
=0.0001$. (a) Linear case. (b) Nonlinear case at $c=0.1$. (c)
shows the real parts of the eigenvalues of $H_{J}$.} \label{fig.4}
\end{figure}

%%%%%%%%%%%%%%%%%%%%%%%%%%%%%%%%%%%%%%%%%%%%%%%%%%%%%%%%%%%%%%%%%%%%%%%%%%%%%%%%%%%%%%%%%%

The above instability mechanism occurs for any smaller nonlinear
perturbation. Let us make some analytic deduction as follows. Note
that the fixed points of equations
(\ref{euqation7})-(\ref{equation 10}) can be
accurately calculated if $c=0:$ $s_{1}^{0}=s_{2}^{0}=\frac{1}{2+\gamma ^{2}}%
, $ $\theta _{1}^{0}=0,$ $\theta _{2}^{0}=\pi $ for $\gamma >0,$ and $\theta
_{1}^{0}=\pi ,$ $\theta _{2}^{0}=0$ for $\gamma <0$. By employing the
perturbation theory using $c$ as small parameter, we can get the fixed
points for small $c:$ $s_{1}^{0}=\frac{1}{2+\gamma ^{2}}-\frac{(1-\gamma
^{2})^{2}}{4(2+\gamma ^{2})}c\gamma ,$ $s_{2}^{0}=\frac{1}{2+\gamma ^{2}}+%
\frac{(1-\gamma ^{2})^{2}}{4(2+\gamma ^{2})}c\gamma ,$ $\theta _{1}^{0}=0,$ $%
\theta _{2}^{0}=\pi $ for nonlinear case. Substituting them into equation (%
\ref{euqation9}), we can obtain the eigenvalues of $H_{J}$ by solving the
following quartic equation:
\begin{equation*}
(64+1280\gamma ^{4})x^{4}+(64+c^{2}+1344\gamma
^{2})x^{2}+(16+c^{2}+352\gamma ^{2})=0
\end{equation*}%
The useful quadratic discriminant is $\Delta =4096\gamma
^{4}-2432c^{2}\gamma ^{2}+(c^{4}-128c^{2}).$ In linear case,
$c=0$, $\Delta =4096\gamma ^{4}$ is always larger than zero, which
means that the solutions for $x$
are pure imaginary, thus the fixed points are stable. For small $c$, $%
\underset{\gamma \to 0}{\lim }\Delta <0,$ the real part of the
solutions $\sim c/16,$ while the imaginary part $\sim \sqrt{2}/2.$
As a result, the fixed point corresponding to mid-level becomes
unstable around $\gamma =0$ for any small nonlinearity, implying
the break down of the adiabatic evolution of states on the
mid-level.

\subsection{Sudden limit ($\protect\alpha\gg 1$)}

The sudden limit corresponds to nonadiabatic LZT. The tunnelling
probability does not relate much to the structure of the levels.
In this limit a weak nonlinearity does not affect the tunnelling
probability, however, a  strong nonlinearity can dramatically
influence the tunnelling dynamics.

In this limit, we can derive the analytical expression of the
tunnelling probabilities using the stationary phase approximation
(SPA). As a demonstration, we concentrate on the mid-level, i.e.
to calculate $\Gamma_{22}$ which equals to
$1-\Gamma_{21}-\Gamma_{23}$. Because of the large sweeping rate
$\alpha $, a quantum state would stay on the initial level most of
the time. Thus the amplitudes $a_{1}$ and$\ a_{3}$ in the
Schr\"{o}dinger equation (1) remain small and $\left| a_{2}\right|
\sim 1$ all the time. A perturbation treatment of the problem
becomes adequate.

We begin with the variable transformation,
\begin{equation}
a_{1}=a_{1}^{^{\prime }}\exp [-i\int_{0}^{t}\left( \frac{\gamma }{2}+\frac{c%
}{4}\left| a_{1}\right| ^{2}\right) dt],
\end{equation}%
\begin{equation}
a_{2}=a_{2}^{^{\prime }}\exp [-i\int_{0}^{t}\left( \frac{c}{4}\left|
a_{2}\right| ^{2}\right) dt],
\end{equation}%
\begin{equation}
a_{3}=a_{3}^{^{\prime }}\exp [-i\int_{0}^{t}\left( -\frac{\gamma }{2}+\frac{c%
}{4}\left| a_{3}\right| ^{2}\right) dt].
\end{equation}%
As a result, the diagonal terms in Hamiltonian are transformed away, and the
evolution equations of $a_{1}^{^{\prime }}$, $a_{2}^{^{\prime }}$, $%
a_{3}^{^{\prime }}$become:
\begin{equation*}
\frac{da_{1}^{^{\prime }}}{dt}=-\frac{v}{2i}a_{2}^{^{\prime }}\exp
[i\int_{0}^{t}\left( \frac{\gamma }{2}+\frac{c}{4}(\left| a_{1}\right|
^{2}-\left| a_{2}\right| ^{2})\right) dt]
\end{equation*}%
\begin{eqnarray*}
\frac{da_{2}^{^{\prime }}}{dt} &=&-\frac{v}{2i}a_{1}^{^{\prime }}\exp
[i\int_{0}^{t}\left( -\frac{\gamma }{2}+\frac{c}{4}(\left| a_{2}\right|
^{2}-\left| a_{1}\right| ^{2})\right) dt] \\
&&-\frac{v}{2i}a_{3}^{^{\prime }}\exp [i\int_{0}^{t}\left( \frac{\gamma }{2}+%
\frac{c}{4}(\left| a_{2}\right| ^{2}-\left| a_{3}\right| ^{2})\right) dt]
\end{eqnarray*}%
\begin{equation*}
\frac{da_{3}^{^{\prime }}}{dt}=-\frac{v}{2i}a_{2}^{^{\prime }}\exp
[i\int_{0}^{t}\left( -\frac{\gamma }{2}+\frac{c}{4}(\left| a_{3}\right|
^{2}-\left| a_{2}\right| ^{2})\right) dt]
\end{equation*}

We need to calculate the above integrals self-consistently. Due to the large
$\alpha $, the nonlinear term in the exponent generally gives a rapid phase
oscillation, which makes the integral small. The dominant contribution comes
from the stationary point $t_{0}$ of the phase around which we have
\begin{equation}
a_{1}^{^{\prime }}=-\frac{v}{2i}\int_{-\infty }^{t}dt\exp
[i\int_{0}^{t}\left( \frac{\gamma }{2}+\frac{3c}{4}\left| a_{1}\right| ^{2}-%
\frac{c}{4}\right) dt].
\end{equation}%
\begin{equation}
\frac{\gamma }{2}+\frac{3c}{4}\left| a_{1}\right| ^{2}-\frac{c}{4}=\alpha
_{1}(t-t_{0}),
\end{equation}%
with
\begin{equation}
\alpha _{1}=\frac{\alpha }{2}+\frac{3c}{4}\left[ \frac{d\left| a_{1}\right|
^{2}}{dt}\right] _{t_{0}}.
\end{equation}%
Since $\left| a_{1}\right| ^{2}=\left| a_{1}^{^{\prime }}\right| ^{2}$ ,then
we have
\begin{equation}
\left| a_{1}\right| ^{2}=\left( \frac{v}{2}\right) ^{2}\left| \int_{-\infty
}^{t}dt\exp \left[ \frac{i}{2}\alpha _{1}\left( t-t_{0}\right) ^{2}\right]
\right| ^{2}
\end{equation}%
This expression can be differentiated and evaluated its result at time $%
t_{0} $. A standard Fresnel integral with the result $\left[ \frac{d\left|
a_{1}\right| ^{2}}{dt}\right] _{t_{0}}=\left( \frac{v}{2}\right) ^{2}\sqrt{%
\frac{\pi }{\alpha _{1}}}$ is obtained. Combining this with the relation
(15), we come to a closed equation for $\alpha _{1}$,
\begin{equation}
\alpha _{1}=\frac{\alpha }{2}+\frac{3c}{4}\left( \frac{v}{2}\right) ^{2}%
\sqrt{\frac{\pi }{\alpha _{1}}}
\end{equation}%
For given $\alpha $, $c$, $v$, $\alpha _{1}$ in the above equation can be
obtained. The tunnelling probability
\begin{equation}
 \Gamma _{23}=\left| a_{1}\right|
_{+\infty }^{2}=\frac{\pi v^{2}}{2\alpha _{1}}
\end{equation}
The alliance of Eq.(20)and (21) gives the analytic expression on
the tunnelling probability $\Gamma_{23}$ in the sudden limit.
Compared with our numerical simulation it shows good agreement at
$c/v<130$,  $c/v> 130$ a clear deviation is observable (Fig.5a).
It is due to the resonance between the "internal field" and the
external field leads to the invalidity of our assumption
$|a_2|\sim 1$, as we show latter.

 Similarly, to calculate $\Gamma_{21}$, we consider following
 equation,
\begin{equation}
a_{3}^{^{\prime }}=-\frac{v}{2i}\int_{-\infty }^{t}dt\exp
[i\int_{0}^{t}\left( -\frac{\gamma }{2}+\frac{3c}{4}\left| a_{3}\right| ^{2}-%
\frac{c}{4}\right) dt]
\end{equation}%
\begin{equation}
-\frac{\gamma }{2}+\frac{3c}{4}\left| a_{3}\right| ^{2}-\frac{c}{4}=\alpha
_{3}(t-t_{0})
\end{equation}%
\begin{equation}
\alpha _{3}=-\frac{\alpha }{2}+\frac{3c}{4}\left( \frac{v}{2}\right) ^{2}%
\sqrt{\frac{\pi }{\left| \alpha _{3}\right| }}
\end{equation}%
\bigskip

Differently, in this case we may have three stationary phase
points that are solutions of  equation (24) when $c<\frac{8%
}{27}\sqrt{\frac{6}{\pi }}\frac{\alpha ^{3/2}}{v^{2}}$, but only one
solution otherwise, as demonstrated in Fig.6. We denote them as $\alpha _{31}$%
, $\alpha _{32}$, $\alpha _{33}$ from smallest to largest. For
small c,  $\alpha _{31}$ is around  $-\alpha /2$ , and the other
two solutions locates at the  two sides of the origin. In this
case,  we simply take $\alpha _{3}=\alpha _{31}+\alpha
_{32}+\alpha _{33}$ .
%   #####################################################################
\begin{figure}[tbh]
\begin{center}
\rotatebox{0}{\resizebox *{9.0cm}{6.0cm} {\includegraphics
{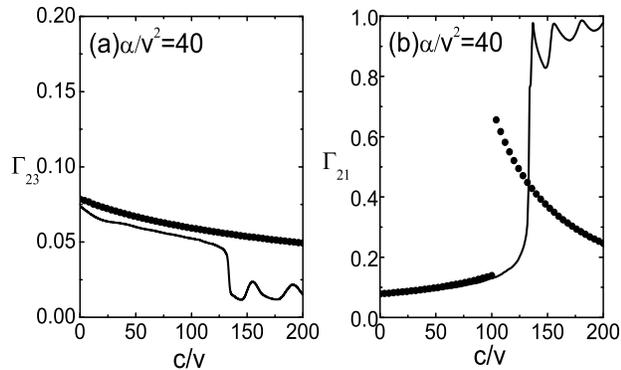}}}
\end{center}
\caption{Comparison between our analytic results using SPA (full
circles and crosses) and the numerical integration of the
Sch\"{o}rdinger equation (1)(solid lines). A cross is used to
denoted the invalidation of SPA } \label{fig.5}
\end{figure}

%   #####################################################################
%   #####################################################################
\begin{figure}[tbh]
\begin{center}
\rotatebox{0}{\resizebox *{9.0cm}{7.0cm} {\includegraphics
{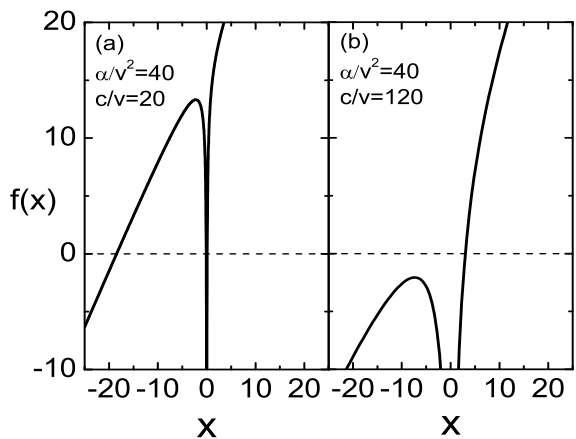}}}
\end{center}
\caption{The plot of function $f(x)=x+\frac{\protect\alpha }{2}-\frac{3cv^{2}%
}{16}\protect\sqrt{\frac{\protect\pi }{|x|}}$} \label{fig.6}
\end{figure}

%   #####################################################################

Then
\begin{equation}
a_{3}^{^{\prime }}=-\frac{v}{2i}\int_{-\infty }^{t}dt\exp [i\int_{0}^{t}%
\frac{\alpha _{3}}{2}(t-t_{0})^{2}dt].
\end{equation}

\bigskip

The tunnelling probability
\begin{equation}
\Gamma _{21}=\left| a_{3}\right| _{+\infty }^{2}=\left| \frac{\pi
v^{2}}{2\alpha _{3}}\right|
\end{equation}
The alliance of the Eq.(24,26) will give the approximate solution
of the $\Gamma_{21}$. Compared with our numerical simulation it
shows a good agreement at $c/v<110$,  whereas for $c/v> 110$ a
clear deviation is observed (Fig.5b).

What happens around $c/v=110$ that leads to the break down of our
stationary phase approximation? The reason is the resonance
between the "internal field" and the external field. Let us recall
the exponent in the integrand of Eq.(22), we find the effective
sweeping rate should be the difference between the change rate of
the "internal field " (i.e., $|a_3|$ ) and the sweeping rate of
the external field. At $c/v=110$, we find the two frequencies
become almost identical, leading to the invalidity of SPA
assumption of rapid phase oscillation. This resonance accompanied
by the bifurcation of the stationary phase points. Crossing
$c/v=110$ we observe the number stationary phase points changes
from three to one, as shown in Fig.6. The resonance breaks the SPA
leading to serious transition from level 2 to level 1,
consequently, at $c/v> 130$, our assumption $|a_2|\sim 1$ become
invalid, and our approximation on the $\Gamma_{23}$ from SPA is no
longer good as shown in Fig.5a.

%%%%%%%%%%%%%%%%%%%%%%%%%%%%%%%%%%%%%%%%%%%%%%%%%%%%%%%%%%%%%%%%%%%%%%%%%%%%%%%%%%%%%%%%%
\begin{figure}[tbh]
\begin{center}
\rotatebox{0}{\resizebox *{10.0cm}{8.0cm} {\includegraphics
{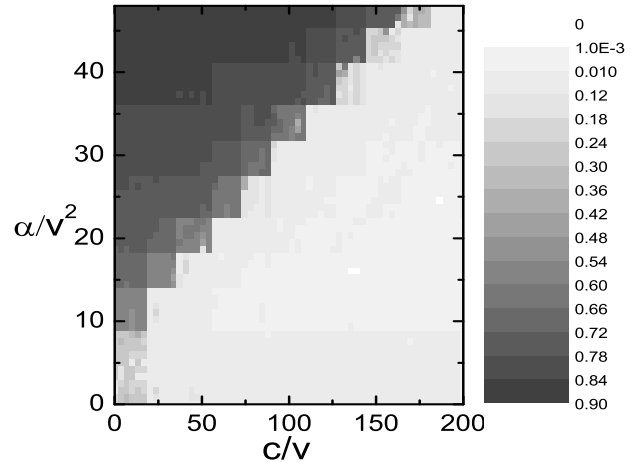}}}
\end{center}
\caption{The contour plot of  tunnelling probability $\Gamma
_{22}$ as the functions of the scaled sweeping rate and
nonlinearity. } \label{fig.7}
\end{figure}
\bigskip

%%%%%%%%%%%%%%%%%%%%%%%%%%%%%%%%%%%%%%%%%%%%%%%%%%%%%%%%%%%%%%%%%%%%%%%%%%%%%%%%%%%%%%%%%%

%%%%%%%%%%%%%%%%%%%%%%%%%%%%%%%%%%%%%%%%%%%%%%%%%%%%%%%%%%%%%%%%%%%%%%%%%%%%%%%%%%%%%%%%%
\begin{figure}[tbh]
\begin{center}
\rotatebox{0}{\resizebox *{8.0cm}{12.0cm} {\includegraphics
{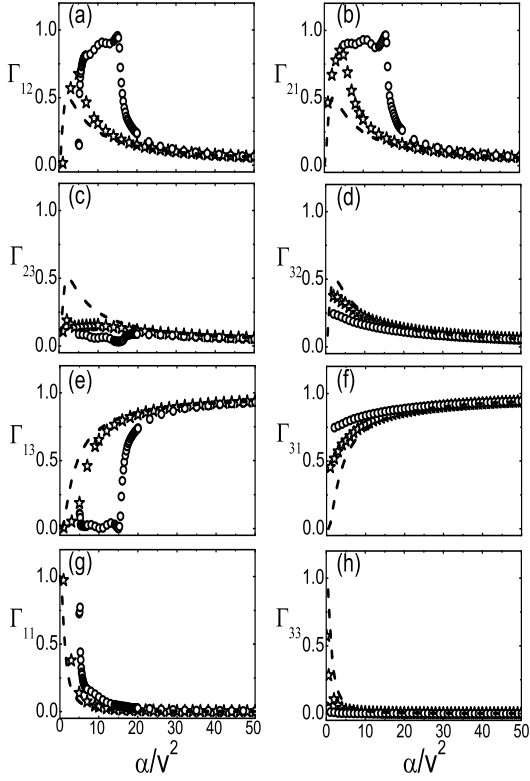}}}
\end{center}
\caption{$\Gamma _{nm}$ as the function of $\protect\alpha $ for
$c=10$ (open pentacles), $c=40$ (open circles) at $v=1.0$. Dashed
line denotes the linear case for comparison. } \label{fig.8}
\end{figure}

%%%%%%%%%%%%%%%%%%%%%%%%%%%%%%%%%%%%%%%%%%%%%%%%%%%%%%%%%%%%%%%%%%%%%%%%%%%%%%%%%%%%%%%%%%

\subsection{General property of the nonlinear tunnelling probability}
The nonlinear tunnelling probability as the function of the two
scaled quantities $\alpha/v^2$ and $c/v$, show many unusual
properties. Taking mid-level tunnelling $\Gamma_{22}$ for example,
we make a large numerical exploration for a wide range of
parameters, to demonstrate the general property of the nonlinear
tunnelling probability in Fig.7. In general, increasing the
sweeping rate will reduce the probability of tunnelling to upper
or lower level and the positive nonlinearity usually suppresses
the probability of the state's staying in the mid-level, because
that the nonlinearity with positive $c$ can be regarded as  a kind
of repulsive potential. This repulsive self-interaction make
particle tend to transition to lower level more easily, and this
transition becomes more serious at the occurrence  of the
resonance between the "internal field" and external field.  The
occurrence of the resonance is clearly exposed by the boundary
between the white regime and dark regime in Fig.7. In the white
regime, due to the resonance, the nonlinearity dramatically
changes the tunnelling probability.

The other issue we want to address is the symmetry. The
nonlinearity makes levels deform and therefore break the symmetry
between upper level and lower level, consequently, the relations
$\Gamma _{21}=\Gamma _{23}=\Gamma _{32}=\Gamma _{12},\Gamma
_{31}=\Gamma _{13},\Gamma _{33}=\Gamma _{11}$ hold in linear case
break in presence of the nonlinearity.
 For our three-level system, the symmetry breaking is clearly exposed by
Fig.8 of   showing the tunnelling probability $\Gamma _{nm}$ as
the functions of $\alpha /v^{2}$ for $c=10$, $c=40$. In the linear
case, we have $\Gamma _{21}=\Gamma _{23}=\Gamma _{32}=\Gamma
 _{12}$, however, with  the presence of the nonlinear,
 $\Gamma_{12}, \Gamma_{21}$ increases whereas the $\Gamma _{23}, \Gamma
 _{32}$ decreases. The similar things happen for the $\Gamma
_{31}, \Gamma _{13}$ and $\Gamma _{33}, \Gamma _{11}$. The above
symmetry breaking may be  observed experimentally\cite{asy}.

\section{Conclusion and Application}

In conclusion, we have made  a comprehensive analysis of the
Landau-Zener tunnelling in  a nonlinear three-level system, both
analytically and numerically. Many novel tunnelling properties are
demonstrated and behind dynamical mechanism is revealed.

Our model can be directly applied to the  triple-well trapped BEC
and to explain the tunnelling dynamics between the traps
\cite{sun,graefe}. In a triple-trap $v(r)$, a BEC is described by
Gross-Pitaevskii equation (GPE) $i\hbar \frac{\partial \Psi
(r,t)}{\partial t} =-\frac{\hbar ^{2}}{2m} \nabla ^{2}\Psi (r,t)+
\left[ v(r)+g_{0}\left| \Psi (r,t)\right| ^{2}\right] \Psi (r,t) $
under the mean-field approximation, where $g_{0}=\frac{4\pi \hbar ^{2}aN}{m}$%
, $m$ the atomic mass and $a$ the scattering length of the
atom-atom interaction. The wave function $\Psi (r,t)$ of GPE is
the superposition of three wave functions describing the
condensate in each trap\cite%
{raghavan}, i.e.,   $ \Psi (r,t)=\psi _{1}(t)\phi _{1}(r)+\psi
_{2}(t)\phi _{2}(r)+\psi _{3}(t)\phi _{3}(r)$. When we study the
tunnelling of three weakly coupled BEC in traps $1$, $2$ and $3$,
the dynamics of the system is described by the nonlinear
Sch\"{o}dinger equation with the Hamiltonian,
\begin{equation}
H=\left(
\begin{array}{ccc}
E_{1}^{0}+c_{1}\left| \psi _{1}\right| ^{2} & -K_{12} & 0 \\
-K_{12} & E_{2}^{0}+c_{2}\left| \psi _{2}\right| ^{2} & -K_{23} \\
0 & -K_{23} & E_{3}^{0}+c_{3}\left| \psi _{3}\right| ^{2}%
\end{array}%
\right) ,  \label{euqation12}
\end{equation}%
where $E_{\alpha }^{0}=\int (\frac{\hbar ^{2}}{2m}\left| \nabla \phi
_{\alpha }\right| ^{2}+v\left( r\right) \left| \phi _{\alpha }\right|
^{2})dr $ $(\alpha =1,2,3)$ is the ground state energy for each trap. $%
c_{\alpha }=\int g_{0}\left| \phi _{\alpha }\right| ^{4}dr$ $(\alpha =1,2,3)$
stands for atom-atom interaction, i.e., nonlinear parameter. $K_{12}=-\int (%
\frac{\hbar ^{2}}{2m}\nabla \phi _{1}\nabla \phi _{2}+v(r)\phi _{1}\phi
_{2})dr$ is the coupling matrix element between trap $1$ and $2$. $%
K_{23}=-\int (\frac{\hbar ^{2}}{2m}\nabla \phi _{2}\nabla \phi
_{3}+v(r)\phi _{2}\phi _{3})dr$ is the coupling matrix element
between trap $2$ and $3$. For simplicity, we only consider the
case that these two coupling matrix elements are the same and
there is no coupling between trap $1$ and $3$, i.e.,
$K_{12}=K_{23}=K$, $K_{13}=0$. The energy bias can be adjusted by
tilting the trapping well and the nonlinearity can be adjusted by
the Feshbach resonance technique. We hope our theory will
stimulate the experiment in the direction.

\bigskip

\section{Acknowledgments}
This work was supported by National Natural Science Foundation of
China (No.10474008,10445005), Science and Technology fund of CAEP,
the National Fundamental Research Programme of China under Grant
No. 2005CB3724503, the National High Technology Research and
Development Program of China (863 Program) international
cooperation program under Grant No.2004AA1Z1220.

\end{document}